\documentclass[review]{elsarticle}

\usepackage{lineno,hyperref}
\usepackage{color}

\modulolinenumbers[5]

\journal{Journal of \LaTeX\ Templates}

\biboptions{sort&compress}

\begin{document}

\begin{frontmatter}

\title{Measuring the branching ratios from the $y^8 {\rm P}_{9/2}$ state to metastable states in europium}


\author{Yuki Miyazawa, Ryotaro Inoue, Keiji Nishida, \\Toshiyuki Hosoya}

\author{Mikio Kozuma\corref{mycorrespondingauthor}}

\cortext[mycorrespondingauthor]{Corresponding author. Tel./fax: +81 3 5734 2451.}
\ead{kozuma@phys.titech.ac.jp}

\address{Department of Physics, Tokyo Institute of Technology, 2-12-1 O-Okayama, Meguro-ku,
	Tokyo 152-8550, Japan}

\begin{abstract}
We measure the branching ratios from the $y^8{\rm P}_{9/2}$ excited state to six metastable states of europium through fluorescence spectroscopy of an atomic beam. 
The sum of the six branching ratios is estimated to be $1.05(2)\times 10^{-3}$. 
This research provides us with insightful information to determine the feasibility of using the $a^8{\rm S}_{7/2} - y^8{\rm P}_{9/2}$ transition in order to implement the Zeeman slowing for europium atoms in the ground state. 
Based on this result, we also propose a scheme for Zeeman slowing and magneto-optical trapping, using a specific metastable state which has a cyclic transition.

\end{abstract}

\begin{keyword}
Europium \sep Laser cooling \sep Branching ratio \sep  Dipolar gas
\PACS 32.70.Fw
\MSC[2010] 70-05
\end{keyword}

\end{frontmatter}


\section{Introduction}
Ultracold atoms with a large magnetic dipole moment $\mu$ are useful to study a new class of strongly correlated physics relying on magnetic dipole-dipole interaction (MDDI). 
Many exciting phenomena based on the long-range and anisotropic nature of the interaction have been studied and pave the way for recent investigations such as supersolid \cite{SupersolidScarola2005, SupersolidYi2007}, quantum magnetism \cite{MagnetismPeter2012, MagnetismPaz2013, Stamper-Kurn2013}, and spontaneous spin textures \cite{Yi2006,Kawaguchi2006,Takahashi2007}.
So far, strongly dipolar atomic species of chromium (Cr, $\mu =6$ Bohr magneton, $\mu_B$), dysprosium (Dy, $\mu =10\,\mu_B$), and erbium (Er, $\mu =7\,\mu_B$) have been brought to quantum degeneracy \cite{Griesmaier2005, Lu2011, Aikawa2012}. 
Rich dipolar phenomena have been observed and studied with these systems, such as the d-wave collapse of the dipolar Bose-Einstein condensate \cite{Lahaye2008}, the deformation of the Fermi surface \cite{Aikawa2014}, and the droplets stabilized by quantum fluctuations \cite{Ferrier-Barbut2016,Chomaz2016}. These experiments are performed under DC magnetic field for fixing spin direction, whereas fascinating phenomena can also be expected under ultra-low magnetic field. The MDDI couples spins and orbital angular momenta, which are believed to produce rich ground-state phases including spin textures and vortices  \cite{Yi2006, Kawaguchi2006, Takahashi2007}. 

Europium (Eu), which has not been laser-cooled yet, is a good candidate for investigating such ground-state phases. It has large dipole moment ($7\,\mu_{\rm B}$) and two stable bosonic isotopes, ${}^{151}$Eu (natural abundance: $48\,\%$) and ${}^{153}$Eu ($52\,\%$), where both the isotopes have the same nuclear spins $I=5/2$. 
Since there exists hyperfine structure in the ground state, a s-wave scattering length can in principle be controlled by using the microwave-induced Feshbach resonance \cite{Papoular2010, 1367-2630-12-8-083031} even under zero magnetic field. 
Note that the bosonic isotopes of Cr, Dy, and Er do not have hyperfine structures in their ground states. 
Holmium (Ho), which has recently been laser-cooled \cite{Miao2014}, is one of the good candidates since it has large dipole moment ($9\,\mu_{\rm B}$) and a stable bosonic isotope (${}^{165}$Ho) with non-zero nuclear spin. 
However, the inelastic loss due to hyperfine exchanging collision \cite{Sesko1989} might be concerned with the investigation of the ground-state phases, since its hyperfine state with maximum magnetic moment does not correspond to the lowest energy one.

Magneto-optical trapping (MOT) is the first building block to bring the Eu atoms into quantum degeneracy. 
A hot effusive oven is one of the most promising sources providing enough atomic flux, since Eu has almost zero vapor pressure at room temperature \cite{Habermann1963}.
The hot atomic beam should be decelerated by the Zeeman slowing technique using a cyclic transition with MHz order of natural linewidth. 
The only available transition for the ground-state Eu is the $a^8{\rm S}_{7/2} - y^8{\rm P}_{9/2}$, where the transition wavelength and the natural linewidth are $460\,\mathrm{nm}$ and $27\,\mathrm{MHz}$, respectively. 
However, the Zeeman slowing of Eu is still challenging, since the values of the branching ratios from the excited state into metastable states are unknown. 
In this paper, we report measurement of the branching ratios from the excited state $y^8{\rm P}_{9/2}$ to six metastable states through spectroscopy using an atomic beam. 
Based on the result, we discuss the possibility of the Zeeman slowing and magneto-optical trapping of Eu.

\section{Optical leak problem}
Europium has complex energy structure as well as Cr, Dy, and Er. 
Effective laser cooling of such atomic species is difficult because of the lack of closed optical transitions. 
When an atom is excited by absorption of a photon, in general, it has many decay channels to corresponding final (or metastable) states. 
For example, the ${}^7{\rm S}_3-{}^7{\rm P}_4$ transition is employed as a transition for the laser cooling of ${}^{52}$Cr \cite{Bell1999} and the excited state ${}^7{\rm P}_4$ has two more relaxation pathways besides that to its ground state ${}^7{\rm S}_3$. 
The probability of leakage out of the cooling transition limits the interaction-time or lifetime because the leakage transition turns off the cooling and trapping mechanism for the atom. 
Such a probability, which is usually referred to as an optical leak, is given by the sum of branching ratios to all final states except for one to the ground state. 
Here the branching ratio from the excited state to a particular final state labeled by $i$ is defined as $\Gamma_{i}/\sum_{i'}\Gamma_{i'}$, where the $\Gamma_{i}$ is the decay rate to the state $i$. 
The first MOT of ${}^{52}$Cr is realized by repumping the leaked atoms back to the cooling transition \cite{Bell1999}, where the optical leak probability is $5\times 10^{-6}$ \cite{Stuhler2001}.
In the case of other dipolar atoms, ${}^{163}$Dy and ${}^{168}$Er also have the optical leak probabilities of $8\times 10^{-6}$ \cite{Lu2010} and $7\times 10^{-6}$ \cite{McClelland2006}, respectively. 
Despite such large optical leak probabilities, MOT of these atomic species is successfully demonstrated without any repumping beams. 
Since they have large magnetic dipole moments, atoms that decay to metastable states are trapped by quadrupole magnetic field used for the MOT and cascade back to their ground states in a few milliseconds.

\begin{figure}[!b]
	\begin{center}
		\includegraphics{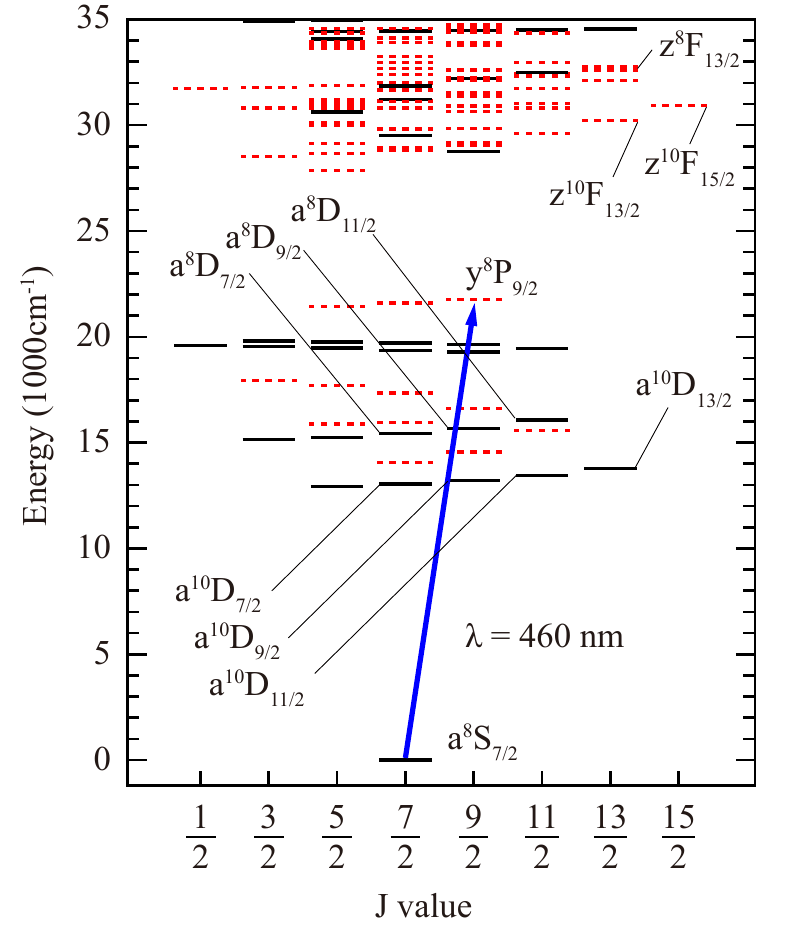}
		\caption{(Color online) Energy levels of europium, showing the $a^8{\rm S}_{7/2} - y^8{\rm P}_{9/2}$ transition. Solid (dotted) horizontal lines indicate odd (even) parity states. {\it J} is the total angular momentum.} 
	\end{center}
\end{figure}

In the case of Eu, the $a^8{\rm S}_{7/2} - y^8{\rm P}_{9/2}$ transition is the only feasible candidate to decelerate an atomic beam. 
Figure 1 shows energy levels of Eu atoms up to $35000\,\mathrm{cm^{-1}}$ \cite{Martin1978, 0067-0049-141-1-255} with the $a^8{\rm S}_{7/2} - y^8{\rm P}_{9/2}$ transition. 
There are eleven electric-dipole transitions from the $y^8{\rm P}_{9/2}$ excited state to metastable states. 
Among the eleven transitions, we measure branching ratios for three intercombination transitions ($y^8{\rm P}_{9/2} - a^{10}{\rm D}_{7/2,\ 9/2,\ 11/2}$, wavelength: $1148\sim1204\,\mathrm{nm}$) and three allowed ones ($y^8{\rm P}_{9/2} - a^{8}{\rm D}_{7/2,\ 9/2,\ 11/2}$, wavelength: $1577\sim1760\,\mathrm{nm}$). 
Note that the other five transitions have much longer wavelengths ($4019\sim4880\,\mathrm{nm}$) and corresponding branching ratios are expected to be smaller than those for the above six transitions, since a decay rate of an electric dipole transition scales as $\propto \lambda^{-3} |d|^2$, where $\lambda$ is a transition wavelength, and $d$ is a transition electric dipole moment. 
When we employ the $a^8{\rm S}_{7/2} - y^8{\rm P}_{9/2}$ as the cooling transition, the lower limit of the optical leak probability is given by the sum of the eleven branching ratios. 

\section{Experimental details}

Figure 2 shows the schematic of our experimental setup. 
The Eu atomic beam is produced by an effusive oven operating at $900\,\mathrm{K}$. 
This temperature provides enough atomic density while keeping the background pressure of the chamber, where the laser and the atomic beam cross each other, is less than $1\times 10^{-4}\,\mathrm{Pa}$ with the atomic beam running. 
The atoms have a mean longitudinal velocity of $360\,\mathrm{m/s}$, and the transverse velocity profile was measured via fluorescence spectroscopy as Gaussian with a standard deviation of $4.4\,\mathrm{m/s}$. 

\begin{figure}[!b]
	\begin{center}
		\includegraphics{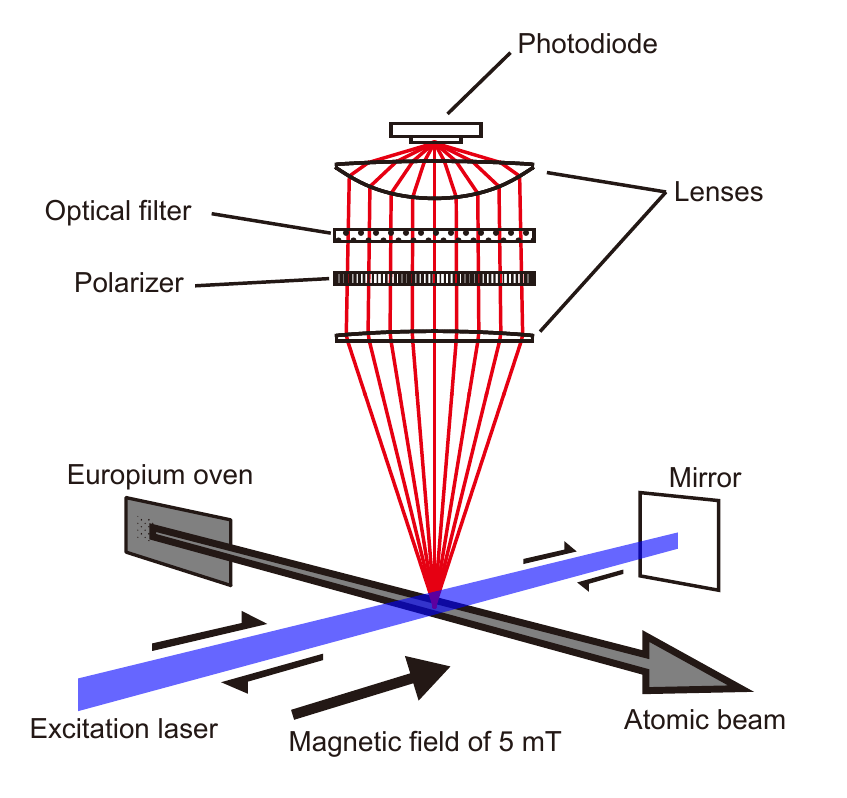}
		\caption{(Color online) Schematic of the experimental setup.} 
	\end{center}
\end{figure}

On-resonant light source for the $a^8{\rm S}_{7/2}\ (F=6) - y^8{\rm P}_{9/2}\ (F=7)$ transition at the wavelength of $460\,\mathrm{nm}$ is prepared by a frequency doubling with an external cavity laser diode at $919\,\mathrm{nm}$. 
The circularly polarized laser beam crosses the atomic beam perpendicularly, where the beam waist and the peak intensity of the laser are $5\,\mathrm{mm}$ and $7\,I_{{\rm sat}}$, respectively. 
Here, $ I_{{\rm sat}}=37\,\mathrm{mW/cm^2}$ denotes the saturation intensity for the transition. 
The excitation laser beam is counterpropagated in order to balance radiation pressure to the atomic beam. 
The fluorescence signal is obtained by sweeping the laser frequency across the resonance.
The flux of the atoms in $F=6$ hyperfine manifold is $8.3\times 10^{11}$/s, which is measured by absorption spectroscopy. 
The atoms are optically pumped to the $m_F=6$ magnetic sublevel at the upper stream of the atomic beam. 

Scattered photons at a wavelength of $460\,\mathrm{nm}$ are detected by a Si-PIN photodiode (Hamamatsu, S3399). 
The photons at infrared wavelengths, which correspond to relaxation channels into metastable states, are also detected by using InGaAs-PIN photodiodes (Hamamatsu, G12180-030A and G12181-230K). 
Fluorescence intensity for each wavelength is measured with combination of seven optical filters, where the transmission spectrum of each filter is evaluated by Fourier transform infrared spectroscopy. 
Figure 3 shows typical photocurrent signals at the wavelength of $460\,\mathrm{nm}$ (a) and at the infrared wavelength (b).  
The branching ratios are estimated by comparing the fluorescence intensities.

Angular distributions of the fluorescence for individual optical transitions are generally anisotropic and different in each, whereas the signals only reflect a part of the fluorescence intensity due to a finite solid angle collection efficiency. 
Additionally, the spatial patterns of the radiation also depend on degree of polarization of atomic spin and/or the excitation laser.
In order to take into account these effects, we measure the fluorescence intensities of two orthogonal polarization components individually with rotating the linear polarizer shown in Fig. 2. 
Here we apply a magnetic field of $5\,\mathrm{mT}$ along the laser beam propagation direction, which we choose as the axis of quantization. 
The radiation patterns of $\pi$ and $\sigma_{\pm}$ polarization components are known to be proportional to $\sin^2\theta$ and $1+\cos^2\theta$, respectively, where $\theta$ is a polar angle with respect to the quantization axis. 
The polarization sensitive detection enables us to eliminate the polarization dependence in our estimation of the branching ratios. 
The branching ratios are also measured with linearly polarized excitation laser, and the results are consistent with the values with circularly-polarized excitation laser (listed in Table 1) within the experimental error; it confirms the validity of the scheme.

\begin{figure}[h]
	\begin{center}
		\includegraphics{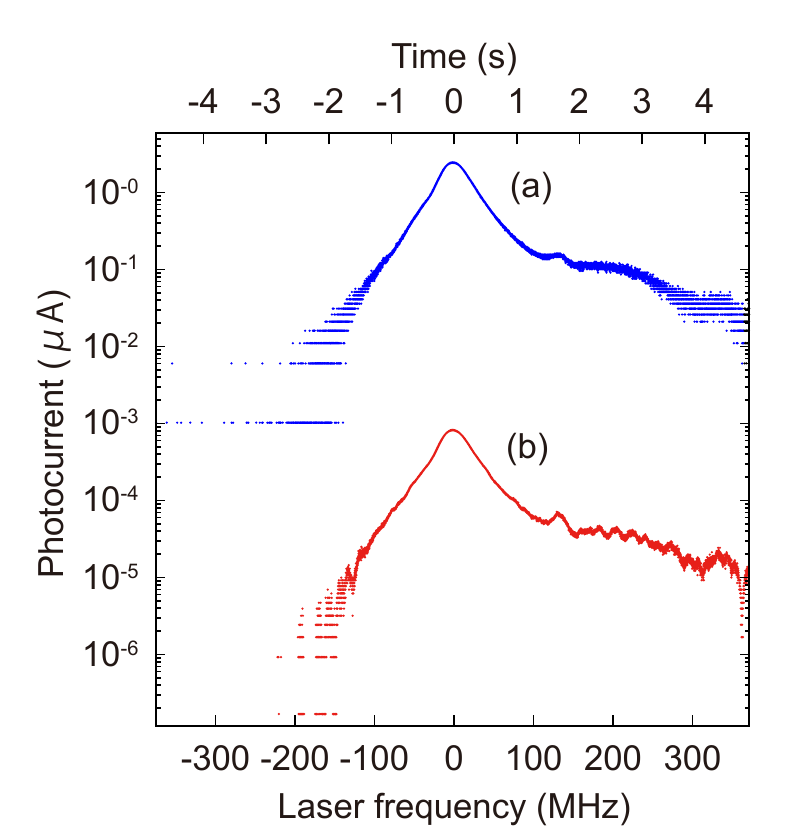}
		\caption{(Color online) Typical photocurrent signals corresponding to  $\sigma_\pm$ polarized fluorescence at $460\,\mathrm{nm}$ (a) and at infrared wavelengths containing all the six transitions (b), plotted on a logarithmic scale.} 
	\end{center}
\end{figure}

\section{Results and discussion}
The measured branching ratios from the $y^8{\rm P}_{9/2}$ excited state to the six metastable states are listed in Table 1. 
Here the number in the parentheses represents the uncertainty combining statistical and systematic errors. 
The primary sources of the errors are uncertainties of the optical filters' transmittances and those of quantum efficiencies for the photodiodes. 
The former comes from the dependence of the transmittance on the angle of the incident fluorescence light.

\begin{table}[htb]
	\begin{center}
		\begin{tabular}{ccc} \hline
			
			\parbox{5em}{\begin{center}Transition\\($y^8{\rm P}_{9/2} - $)\end{center}} &
			\parbox{5em}{\begin{center}Transition\\ wavelength\end{center}} &
			\parbox{5em}{\begin{center}Branching\\ratio\end{center}} \\  \hline
			$a^{10}{\rm D}_{7/2}$ & $1148\,\mathrm{nm}$ &  $ 0.32(6)\times 10^{-4} $   \\
			$a^{10}{\rm D}_{9/2}$ & $1171\,\mathrm{nm}$ &  $ 1.08(6)\times 10^{-4} $   \\
			$a^{10}{\rm D}_{11/2}$ & $1204\,\mathrm{nm}$ &  $ 1.78(6)\times 10^{-4} $  \\ 
			$a^8{\rm D}_{7/2}$ & $1577\,\mathrm{nm}$ &  $ 0.24(5)\times 10^{-4} $      \\
			$a^8{\rm D}_{9/2}$ & $1644\,\mathrm{nm}$ &   $ 1.33(6)\times 10^{-4} $     \\
			$a^8{\rm D}_{11/2}$ & $1760\,\mathrm{nm}$ &   $ 5.72(10)\times 10^{-4} $   \\ \hline \hline
			\multicolumn{2}{c}{Total} & $ 1.05(2)\times 10^{-3} $ \\ \hline
		\end{tabular}
		\caption{Estimated branching ratios.}
	\end{center}
\end{table}

The sum of the six branching ratios is $1.05(2)\times 10^{-3}$, which is two orders of magnitude lager than the optical leak probabilities reported in the experiments using Cr \cite{Stuhler2001}, Dy \cite{Lu2010}, and Er \cite{McClelland2006}. 
As can be seen in Table 1, at least six repumping beams are required to reduce the optical leak probability to the same extent as the values for Cr, Dy and Er. 
Since there exist five more decay channels, eleven repumping lights are necessary to completely plug the all optical leaks.

Considering such large branching ratios, we here discuss the possibility to laser-cool Eu in the specific metastable state having a cyclic transition. 
The transition from the $a^{10}{\rm D}_{13/2}$ metastable state to the $z^{10}{\rm F}_{15/2}$ is completely closed in electric-dipole approximation (see Fig. 1), where the wavelength and the natural linewidth are $583\,\mathrm{nm}$ and $8.3\,\mathrm{MHz}$, respectively \cite{0067-0049-141-1-255}. 
The transition is the cyclic one with MHz order of natural linewidth and can thus be applicable to conventional Zeeman slowing and also magneto-optical trapping techniques. 
In order to prepare atoms in the $a^{10}{\rm D}_{13/2}$ metastable state, we make use of the large branching ratios. 
One can start from pumping the ground state of atoms to the $a^8{\rm D}_{11/2}$ metastable state by using the $a^8{\rm S}_{7/2} - y^8{\rm P}_{9/2}$ transition. 
Based on the branching ratios estimated by our measurements, the transfer efficiency of $55\,\%$ can be obtained. Through successive optical pumping using the $a^8{\rm D}_{11/2} - z^8{\rm F}_{13/2}$ transition (wavelength: $599\,\mathrm{nm}$), $92\,\%$ of atoms in the $a^8{\rm D}_{11/2}$ state can be pumped to the $a^{10}{\rm D}_{13/2}$ state \cite{0067-0049-141-1-255}. 
The total transfer efficiency from the ground state to the $a^{10}{\rm D}_{13/2}$ metastable can thus be estimated to be $51\,\%$. To perform the dipolar gas experiments with the condensate of Eu atoms, atoms have to be pumped back to the ground state after the Zeeman slowing and MOT processes using the $a^{10}{\rm D}_{13/2} - z^{10}{\rm F}_{15/2}$ transition. 
By using the $a^{10}{\rm D}_{13/2} - z^{10}{\rm F}_{13/2}$ transition (wavelength: $608\,\mathrm{nm}$), $94\,\%$ atoms in the $a^{10}{\rm D}_{13/2}$ can be transferred to the $a^{10}{\rm D}_{11/2}$ metastable state \cite{0067-0049-141-1-255}. 
Through successive pumping with the $a^{10}{\rm D}_{11/2} - y^8{\rm P}_{9/2}$ transition (wavelength: $1204\,\mathrm{nm}$), most of the atoms in the $a^{10}{\rm D}_{11/2}$ state can be transferred to the ground state. 
Laser cooled atoms in the $a^{10}{\rm D}_{13/2}$ metastable state can thus be pumped back to the ground state with the efficiency of $95\,\%$.

\section{Conclusion}
We have spectroscopically measured the branching ratios from the $y^8{\rm P}_{9/2}$ excited state to the six metastable states. 
The sum of the six branching ratios are estimated to be $1.05(2)\times 10^{-3}$, which indicates that laser-cooling of Eu atoms in the ground state requires at least six, maybe eleven repumping light sources. 
As a simpler alternative to preparing many repumping lights, we have also proposed the new scheme which utilizes the metastable state $a^{10}{\rm D}_{13/2}$. 
The atoms can be efficiently pumped via all-optical way to the metastable state with the help of the large branching ratio. 
The $a^{10}{\rm D}_{13/2} - z^{10}{\rm F}_{15/2}$ transition enables us to operate the Zeeman slowing and MOT of the metastable Eu, and the laser-cooled atoms can also be pumped back to their ground state.

\section*{Acknowledgments}
We sincerely thank B. P. Das for his valuable comments and many stimulating discussions. This work was supported by a Grant-in-Aid for Challenging Exploratory Research; the Murata Science foundation, and the Research Foundation for Opto-Science and Technology.


\bibliography{mybibfile}

\end{document}